\documentclass[12pt]{article}

\usepackage{float}
\usepackage{amsfonts}
\usepackage{bm}
\usepackage{amsmath}
\usepackage{amscd}
\usepackage{amssymb,lscape}

\newcommand{\bmat}{\left(\begin{array}}
\newcommand{\emat}{\end{array}\right)}
\def\gtrsim{\mathrel{\raise.3ex\hbox{$>$\kern-.75em\lower1ex\hbox{$\sim$}}
}
}

\def\vec#1{\textbf{\emph{#1}}}

\def\ap{\alpha^{\prime}}

\def\-{\hphantom{-}}

\def\s2{\frac{1}{\sqrt2}}

\def\beq{\begin{equation}}
\def\eeq{\end{equation}}
\def\beqa{\begin{eqnarray}}
\def\eeqa{\end{eqnarray}}

\def\mg{m_{3/2}}
\def\mg2{m^2_{3/2}}

\def\Dsl{\,\raise.15ex\hbox{/}\mkern-13.5mu D} 

\def\be{\begin{equation}}
\def\ee{\end{equation}}
\def\bea{\begin{eqnarray}}
\def\eea{\end{eqnarray}}

\newcommand{\nn}{\nonumber}

\def\sqrtap{\sqrt{\alpha '}}

\makeatletter
\@addtoreset{equation}{section}
\makeatother

\begin{document}
\pagestyle{plain}
\begin{titlepage}
\begin{center}
  \LARGE{ Gauge symmetry enhancing-breaking from a  Double Field Theory 
perspective \\}
\large{\bf  G. Aldazabal${}^{a,b}$, E. Andr\'es$^{a}$
, Mart\' in
Mayo${}^{a}$, J. A. Rosabal${}^c$
 \\}
\small{ ${}^a$ {\em G. F\'isica CAB-CNEA and CONICET, }\\{\em Centro At\'omico 
Bariloche, Av. Bustillo 9500, Bariloche, 
Argentina.}\\ 
${}^b${\em Instituto Balseiro} \\[-0.3em]
{\em Centro At\'omico 
Bariloche, Av. Bustillo 9500,  Bariloche, 
Argentina.}\\[-0.3em]}
\small{ ${}^c${\em 
 B.W. Lee Center for Fields, Gravity {\rm \&} Strings \\
Institute for Basic Sciences, Daejeon 34047 \rm KOREA
.}\\}
\end{center}

\noindent
 {\bf Abstract}:
Gauge symmetry enhancing, at specific points of the compactification space, is a   
distinguished feature of string theory. In this work we discuss the breaking of 
such symmetries with tools provided by 
Double Field Theory (DFT). As a main guiding example we discuss the bosonic 
string compactified on a circle where, at the self-dual radio the generic 
$U(1)\times U(1)$  gauge symmetry becomes enhanced to $SU(2)\times SU(2)$. We 
show that the enhancing-breaking of the gauge symmetry can be understood through a 
dependence of gauge structure constants (fluxes in DFT) on moduli. 
This dependence, in DFT description,  is encoded in the generalized tangent 
frame of the double space.
The explicit T-duality invariant formulation provided by DFT proves 
to be a helpful ingredient.
The link with string theory results is discussed and generalizations to generic 
tori compactifications are  addressed.

\vspace{.5cm}
\today


\end{titlepage}


\begin{small}
\tableofcontents
\end{small}

\newpage\section{Introduction}
\label{sec:Introduction}

The extended nature of strings is responsible for  several amazing phenomena 
that are not conceivable from a field theory of point particles. When moving  on 
compact space, besides the expected states associated to KK 
compact momenta, a  string can  wind around non-contractible cycles leading to 
the so-called winding states, with the winding number being an integer counting the 
number of times that the cycle is wrapped by the string. 
Quantum states are thus labelled by specific values of KK momenta 
and windings.

The  interplay among winding and momentum modes underlies 
T-duality, a genuine stringy feature. Such interplay  manifests itself 
by connecting the 
physics of strings defined on geometrically very different 
backgrounds. At specific points of  moduli of the compact space, states in 
some combinations of windings and momenta  become massless and can give rise to 
enhanced gauge symmetries (see for instance 
\cite{Narain:1985jj,Giveon:1994fu}). 
The simplest example is provided by the compactification of the bosonic string 
on a circle of radio $R$. The resulting theory, which contains a $U(1)\times 
U(1)$ gauge group,  is equivalent to a string compactified on a circle of 
radio $\tilde R=\frac{\ap}{R}$ (where $\ap$ is the string constant) if momenta and 
winding are exchanged. At the self-dual point $R=\tilde R= \sqrt{\ap}$ the 
gauge symmetry is enhanced to $SU(2)\times SU(2)$.

When the compact space is a $r$ dimensional torus $T^r$, characterized by 
some background moduli (internal metric and anti-symmetric fields),  T-duality 
implies that backgrounds related by  the non-compact group $O(r,r,{\mathbb  
Z})$ 
are physically equivalent. Generically a richer structure of points of gauge 
enhancing appear. 
 
Recall that, from the world sheet point of view,  states are created by vertex 
operators involving both coordinates associated with momentum excitations and 
dual coordinates associated to  winding excitations or, equivalently,  to left 
(L)
and right (R) moving coordinates. For  generic values of moduli an Abelian 
symmetry $U(1)_L^r\times U(1)_R ^r$ appears. However, at specific points, the 
symmetry  becomes enhanced to  a gauge symmetry $G_L\times G_R$ where  $G_{L(R)}$ 
are  non-Abelian gauge groups of rank $r$. For 
example, in a two torus $T^2$, a generic $(U(1)\times U(1))_L \times 
(U(1)\times U(1))_R$ is enhanced to $ SU(3)_L\times 
SU(3)_R $ or $(SU(2)\times SU(2))_L\times (SU(2)\times SU(2))_R $ etc. at 
different points.

Let us sketch, as motivation of our work,  the case of circle compactification  
at self-dual point\footnote{Details are presented in next section.}.  The 
effective action in $d$ dimensional space, computed from string theory 3-point
amplitudes \cite{agimnr} reads
\bea \label{effectiveSD}\nn
S&=& \frac{1}{2\kappa_d^2}\int d^dx\sqrt{g}e^{-2\varphi}
\left(
{\cal
R}+4\partial^\mu\varphi\partial_\mu\varphi-\frac1{12}H_{\mu\nu\rho}H^{\mu\nu\rho
}\right) \\\nn &-&\frac 18 \left(\delta_{ij}{F}^{i\mu\nu}{F}^j_{\mu\nu} + 
\delta_{ij}{\bar F}^{i\mu\nu} {\bar F}^j_{\mu\nu} -\frac 12 g_d\sqrt{\ap} M_{ij} 
F^i_{\mu\nu} \bar F^{j\mu\nu }\right) \nn\\
 &-&
 D_\mu M_{ij}D_\nu M^{ij}g^{\mu\nu}
 +\frac{16g_d}{\sqrt{\ap}} \det M  + {\cal
 O}(M^4),
 \label{su2action}
\eea
where  the first row contains the universal gravity contribution, the second 
one contains the gauge field strength for the vector fields of $ SU(2)_{L}$ and 
$ SU(2)_{R}$ (that we denote here as $ A_{L\mu}^i, A_{R\mu}^i$ respectively). 
$M_{ij}$ is the  matrix of scalars living in the $(\bf 3, \bf 3)$ 
representation. $  D_\mu M_{ij}=\partial_\mu M_{ij}+
g_d f^k{}_{li} A^l_\mu
     M^{{}}_{kj} + g_df^k{}_{lj} A^l_\mu M^{{}}_{ik}
\,$
are the usual covariant derivatives  and   $f_{ijk}=-\bar 
f_{ijk}\propto \epsilon_{ijk}$ ($i,j=1,2,3$) are the  structure 
constants where $\epsilon_{ijk}$ is the usual Levi-Civita completely 
antisymmetric tensor and  $g_d=\kappa_{d}\sqrt{\frac{2}{\ap}}  $.

Interestingly enough, this action can be embedded into an $O(d+3,d+3)$ 
framework.
This is discussed in Ref\cite{agimnr} (and briefly reviewed 
below) where it was observed that the spectrum of the 
bosonic string has $(d+3)^2$ massless states: $d^2$  from $g_{\mu\nu}$ and $B_{\mu \nu}$,
$6d$  from the vector states and $9$  the scalar states. 
The number of degrees of freedom precisely agrees with the dimension of the 
coset
\begin{equation}
\frac{O(d+3,d+3)}{O(d+3)\times O(d+3)}
\label{cosett1}
\end{equation}
that counts the number of degrees of freedom in the DFT formulation with
symmetry $O(d+3,d+3)$.

In general a DFT action with $O(D,D)$ symmetry with  $D=d+n$,   can be 
written as 
\bea \nn
S_{eff}&=& \frac{1}{2\kappa_d^2}\int d^dx\sqrt{g}e^{-2\varphi}
\left[
     {\cal R}+4\partial^\mu\varphi\partial_\mu\varphi-\frac1{12}H_{\mu\nu\rho}
     H^{\mu\nu\rho} \right. \\\nn 
     &&-\frac 18{\cal H}^{{}}_{IJ}{ F}^{I\mu\nu}
{F}^J_{\mu\nu}
+\frac 18 (D_\mu {\cal H})_{IJ}
(D^\mu {\cal H})^{IJ}\\\nn
& &-\frac 1{12}f_{IJ}{}^Kf_{LM}{}^N \left( {\cal 
H}^{IL}{\cal 
H}^{JM}{\cal H}_{KN} \right. 
- 3\, {\cal H}^{IL} \eta^{JM} \eta_{KN}\\ & & \left. + 2 \, \eta^{IL} 
\eta^{JM} \eta_{KN}
) 
- \Lambda
\right].\label{dftaction}
\eea
after a generalized Scherk-Schwarz\cite{effective}  like $n$ dimensional  
compactification. 
In this expression ${\cal H}_{IJ}$ with $I, J= 1, \dots, 2n$
is the, so-called,  generalized metric containing the scalar fields coming from the
internal components of the $n$-dimensional metric and $B$-field.
${\cal R}$ is the $d$-dimensional Ricci scalar and the field strengths
$F_{\mu\nu}^A$ and
$H_{\mu\nu\rho}$ are
\bea \label{FHDFT}
F^I & = & d A^I + \frac{1}{\sqrt{2}}f_{JK}{}^I A^J \wedge A^K \nn
\\
H   & = & d B + F^I \wedge A_I  ,
  \eea

The covariant derivative of the scalars is
  \be \label{DH}
  (D_\mu {\cal H})_{IJ}=(\partial_\mu {\cal H})_{IJ}+
\frac{1}{\sqrt{2}}f^K{}_{LI} A^L_\mu
     {\cal H}^{{}}_{KJ} + \frac{1}{\sqrt{2}}f^K{}_{LJ} A^L_\mu{\cal H}^{{}}_{IK}
\,
  \ee
 The structure constants $ f_{NLI}= \eta_{NK}f^K{}_{LI}$ are 
completely antisymmetric and  $ \eta_{NK}$ is the  $O(n,n)$ metric
  \be \label{etam}
\eta^{PQ}=\begin{pmatrix} 1_n & 0 \\ 0 & -1_n \end{pmatrix}.
\ee
 
 In our example   $D=d+3$,  thus $I=1,\dots 6$. The  gauge fields are 
 $A^I_\mu=( A^i_{L\mu}, -A^i_{R\mu})$   and the structure constant 
splits into
 \be
f_{IJ}{}^K=\begin{cases}  (\frac{2}{\ap})^{\frac12}\epsilon_
{ijk}\\
\nonumber 
-(\frac{2}{\ap})^{\frac12}\bar{\epsilon}_{ijk} \end{cases} 
\label{su2su2cc} .
\ee
After expanding around a fixed background the internal generalized metric ${\cal H}_{IJ}$ can be written as 
 \beq \label{Hint}
{\cal H}\simeq \begin{pmatrix}
    1_3 & M\\
  M^T & 1_3
\end{pmatrix}= I+\begin{pmatrix}
    0 & M\\
  M^T & 0
\end{pmatrix}\eeq

By replacing above expressions into the action (\ref{dftaction}), and after 
absorbing constants into the fields,  the $SU(2)\times 
SU(2)$ theory given in (\ref{effectiveSD}) is reproduced.
Of course, any reference to DFT could be omitted and just present the above 
(\ref{dftaction}) action as an interesting way of writing the original 
expression.

It is worth looking at the term containing the derivatives of scalar fields. 
Since the metric ${\cal H}=I+\dots $ contains a constant term, the identity, 
the action could have a contribution, 

  \bea |D_\mu {\cal H}|^2&\equiv& (\dots +  \frac{1}{\sqrt{2}}f^K{}_{LI} 
A^L_\mu
     {\delta }^{{}}_{KJ} + \frac{1}{\sqrt{2}}f^K{}_{LJ} A^L_\mu{\delta
}^{{}}_{IK})^2\\\nn
&=&\dots +  \frac{1}{2}([f^J{}_{LI} 
+ f^I{}_{LJ}] A^L_\mu)^2.
\label{massterm}
  \eea
  Namely, a potential ``mass term'' for the vector bosons.
  
  Moreover, by splitting the $O(n,n)$ indices into Left and Right indices, that 
we denote as  $A=(a,\hat a)$ and by using that $ f_{ABC}=\eta_{AA'} 
f^{A'}{}_{BC}$ is 
completely antisymmetric, the above term can be recast as
  \begin{equation}
 \begin{aligned}
 \big( f_{ABC} A^{B}_{\mu} \delta ^{C}_{D} + f_{DBC} A^{B}_{\mu}\delta^{C}_{A}  
\big)^{2}&\sim A^{B}_{\mu}A^{E\,\mu} f_{ABC}f_{DEF}\big( \eta^{AD}\eta^{CF} - 
\delta^{AD}\delta^{CF} \big)\\
 &\sim A^{B}_{\mu}A^{E\,\mu} f_{aB\hat{c}}f_{aE\hat{c}}
 \end{aligned}
 \end{equation}
  where a sum over repeated indices is understood.
  
Since in our example $f_{IJK}=(f_{ijk},\bar f_{\hat i\hat  j\hat k})$ the 
first three ``Left'' indices do not mix with the
last three ``Right'' ones,   such terms vanish. 

However, we could envisage a situation where Left and Right indices do mix. In 
fact, this is what we expect from string theory when we move away from the self 
dual point. Vertex operators that at the dual point depend only on Left 
coordinates (or Right coordinates) acquire a mixed dependence and the group 
breaks down to $U(1)_L\times U(1)_R$ (in the circle example).
From this observation we could imagine a description of the symmetry breaking 
where the structure constants have a dependence on the moduli, namely 
$f_{IJ}{}^K (R)$, such that for the dual point  $R=\tilde R=\sqrt{\ap}$ 
Left and Right indices do not mix but generically do, away from the fixed point.
Let us propose, out of the blue, the following constants
\bea\nonumber
f_{ij}{}^k&=&\epsilon_{ijk} \frac{1}{\sqrt2 \ap}m_+=-f_{\hat i \hat j}{}^{\hat 
k}\\
f_{12}{}^{\hat 
3}&=&f_{\hat 1 
\hat 2}{}^{3}=f_{1\hat 3}{}^{2}=f_{\hat 13}{}^{\hat 
2}=f_{3\hat 2}{}^{\hat 1}=f_{\hat 32}{}^{1}
=-\frac{1}{\sqrt2 \ap}m_-
\label{structureconstantsr}
\eea
with $\hat i\equiv i+3$ and 
\begin{equation}
 m_{\pm} = \frac{1}{R}\pm \frac{1}{\tilde R}
 \label{m+-}
\end{equation}

If we go back to equation (\ref{massterm}) and replace above flux 
values we find 
\begin{equation}
 A^{B}_{\mu}A^{E\,\mu} f_{aB\hat{c}}f_{aE\hat{c}}\propto m_-^2 
({A^+_{\mu}})^2+ m_-^2 ({A^-_{\mu}})^2
\end{equation}
with $A^{\pm}_\mu =A^1_\mu \pm iA^2_\mu$ acquiring a mass $m_-$ whereas 
$A^3$ and $\bar A^3$ remain massless, indicating that the gauge group is 
spontaneously broken to $U(1)_L\times 
U(1)_R$.
Moreover, by looking at the couplings 
$\eta^{IN}\eta^{JM}\partial_\mu {\cal 
H}_{IJ}f^K{}_{LI}\delta_{NS}A^L_\mu$ we notice that there is a coupling 
\begin{equation}
 \propto m_-\partial_\mu M_{\pm,\hat 3}A^{\mp}_\mu
\end{equation}
with $M_{\pm,\hat 3}=M_{1,\hat 3}+iM_{2,\hat 3}$, 
identifying $M_{\pm,\hat 3}$ as would be Goldstone bosons

In fact, by  replacing the proposed structure constants in the action 
(\ref{dftaction}) and after some redefinitions it can be shown that the full 
string theory effective action \cite{agimnr}, computed  away from the self-dual 
point  (and keeping slightly massive terms), is reproduced.

Interestingly enough, the  structure constants in the action (\ref{dftaction}) 
can be 
understood from a DFT perspective. In  generalized  Scherk-Schwarz reduction 
of DFT \cite{effective, reviewamn}
they appear as  the generalized fluxes of the  algebra 
associated to a generalized vielbein on a doubled internal space. Indeed, it was 
shown in \cite{agimnr} that such a generalized frame can be explicitly 
constructed  to account for the description of the circle compactification at 
the self-dual point.
In the following sections we indicate how to generalize this frame  in 
order 
to provide a description valid also (slightly) away from the point of 
enhancing. The 
constants presented in (\ref{structureconstantsr}) are then obtained as 
generalized fluxes from this frame. It is worth emphasizing that, therefore,  
the resulting DFT construction involves, besides massless states,  massive 
states that  become massless at the fixed point. The breaking is 
not  achieved by giving vacuum expectation values (vev's) to scalar fields. 
We also indicate how to extend the construction  to 
toroidal compactifications in more dimensions and provide some examples for 
the $T^2$ case.
An interpretation from the string theory 
point of view is  provided as well. 

In Section \ref{sec:DFT} we review  the basic ideas of  Double 
Field Theory, with special emphasis on the symmetry enhancing situation and  by 
highlighting the ingredients needed in our construction.
In particular we discuss how to extend the frame description, away from points 
of enhancing, from the circle compactification example.

In Section \ref{sec:extensions} we discuss how to extend the DFT 
construction to describe the enhancing-breaking of gauge symmetries at 
different points of $m$ dimensional toroidal compactification. The structure of 
the gauge groups associated to fixed points is known to be of the form 
$G_L\times G_R$ where  $G_L(R)$ are  non 
Abelian gauge groups of rank $m$.

Concluding remarks and a brief  outlook are presented in Section
\ref{sec:Summary and outlook}.

\section{DFT and enhanced gauge symmetries}
\label{sec:DFT}

Generalized Complex Geometry (GCG) \cite{Hitchin, GMPW, WaldramOdd} and
Double Field Theory (DFT) \cite{hz} are proposals that aim at integrating 
 T-duality as a geometric symmetry. In DFT the presence of windings, an 
essential ingredient of T-duality, is achieved by introducing new coordinates 
associated to the winding numbers. Thus in DFT fields depend on a double set of 
coordinates. This idea,  first proposed in  \cite{duff,tseytlin,siegel}, 
received 
new impulse in recent years \cite{hk, Jeon:2011cn, berwest} (see 
\cite{reviewamn, reviews} for some reviews on 
the subject and references 
therein). Generically, these double field theories are constrained theories 
since some consistency conditions must be satisfied to ensure closure of 
generalized diffeomorphism algebra.
A quite restrictive condition, the so called  section condition (or strong 
constraint),  ensures consistency  at the price of eliminating half of 
the coordinates and, therefore, abandoning the original motivation. However, it 
is worth emphasising that this 
constrained  DFT, which in this case essentially coincides with  GCG, 
still provides an interesting description for
understanding underlying symmetries and  stringy features (for instance 
$\alpha '$ corrections have been recently incorporated
\cite{alphap,alphap2} in these formulations).
An alternative constraint is provided by generalized Scherk-Schwarz 
like compactifications \cite{ss} of DFT \cite{effective}.
These compactifications contain the generic gaugings
of gauged supergravity theories \cite{Samtleben:2008pe, Trigiante:2016mnt} 
allowing for a geometric  interpretation
of all of them. In this  framework, the double coordinates enter in a very 
particular way through the twist matrix. Constant gaugings are computed from 
this matrix and, generically, closure of the algebra is ensure  if 
these gaugings satisfy some quadratic constraints \cite{gm} with no need of a 
strong constraint requirement. A generalization of this formalism  was 
proposed 
in \cite{agimnr} in 
order to account for the description of gauge enhancing. The proposal of 
\cite{agimnr}, 
discussed for the example of circle compactification on $D=d+1$ and inspired in 
the relation with the coset (\ref{cosett1}),  requires to introduce an 
extended tangent space with $d+1\rightarrow d+1+2$. However, the 
``physical space'' of DFT is still a double circle. The  frame vectors do 
depend  on both circle compact coordinates $y$ and its dual $\tilde y$ thus 
being truly non-geometric. 
We strongly rely on these results below in order to 
describe the breaking of enhanced symmetries when moduli do slide slightly away 
from the fixed points. In this process  slightly massive states (those 
that become massless at the self 
dual points), with Kaluza Klein(KK) momenta and windings, are expected to 
appear and are associated to an unpaired number  
 ($N\ne \bar{N}$) of Left and Right moving oscillators.  We keep these states 
and 
disregard other 
massive states contributions. Such a situation was addressed in 
\cite{agimnr}  from the point of view of string theory. Here we show that 
DFT 
is able to capture it.
Let us recall that massive states, including winding and momenta,  were also 
considered in DFT context in \cite{Aldazabal:2016yih} as  generalized 
Kaluza Klein modes  but with equal number of L and R oscillators .

In what follows we, briefly,  review some basic features of GCG and/or
DFT. The theory is defined on a generalized tangent bundle which locally is
$TM\oplus T^{*}M$ and whose sections, the generalized vectors $V$, are direct 
sums of vectors $v$ plus one forms $\xi$, $V=v+\xi $.

A generalized frame $E_A$ on this bundle is a set of linearly independent
generalized vectors that belong to the vector space of representations of the
group $G=O(D,D)$. It parametrizes the coset $G/G_c$,
the quotient being over the maximal compact subgroup of $G$. A  Lorentz 
signature is 
assumed on the $D$-dimensional space-time, i.e. $G_c=O(1,D-1)\times O(D-1,1)$.
In DFT, this generalized tangent bundle  is locally  parametrized with a double 
set of 
coordinates,  
${\mathbb X}^{\cal 
M}=(
 x^{\hat\mu},\tilde x_{\hat\mu},)$, defined in the fundamental
representation of $O(D,D)$. Here ${\cal M}=0, \cdots, 2D$ and $\hat\mu=0, 
\cdots, 
D-1$.

A natural pairing between generalized vectors is defined by
\beq
V_1 \cdot V_2= \iota_{v_1} \xi_2 + \iota_{v_2} \xi_1=\eta(V_1,
V_2)=V_1^{\cal M} \eta_{\cal M \cal N} V_2^{\cal N}\, ,\label{product} \
\eeq
where the $O(D,D)$ metric $ \eta_{\cal M \cal N}$  has the 
following
off-diagonal form
\beq
 \eta_{\cal M \cal N}=\begin{pmatrix} 0 & 1_D \\ 1_D & 0 \end{pmatrix}\ 
,
\label{etamn}
\eeq
where $1_D$ is the $D\times D$ identity matrix. Note that $ \eta_{\cal M \cal 
N}$ is
invariant under ordinary diffeomorphisms. Defining $\eta_{\cal 
AB}=\eta(E_{\cal 
A},E_{\cal 
B})$
where ${\cal 
A},{\cal 
B} =0,1,..,2D$ are frame indices it results that $\eta_{\cal 
AB}$ has the 
same 
numerical form as (\ref{etamn}).

A generalized metric can be constructed as $
{\cal H}_{\cal M \cal N} = E^{\cal 
A}{}_{\cal M}\ S_{{\cal A
B}}\ E^{ {\cal 
B}}{}_{ \cal 
N}\, ,
$ 
where ${\cal S}^{\cal A
B}=\text{diag}(s^{ab},s_{ab})$,
$s_{ab}$ being the Minkowski metric.

The generalized metric can be parametrized as 

\begin{equation}
\mathcal{H}_{\cal MN}({\mathbb X})\,  =\, 
\begin{pmatrix}g^{-1} & -g^{-1}\, B\\
B\, g^{-1}& g-B\, g^{-1}\,B\end{pmatrix}\, ,\label{gm}
\end{equation}
satisfying 
\begin{equation}
\mathcal{H}_{\cal MP}\,\eta^{\cal PQ}\,\mathcal{H}_{\cal QN}\, =\, \eta_{\cal 
MN} \, .
\end{equation}
where $g_{\hat\mu \hat\nu}({\mathbb X}), B_{\hat\mu \hat\nu}({\mathbb X})$ are a 
symmetric and an anti-symmetric tensor, respectively.

The generalized vectors transform under generalized diffeomorphisms as
\bea \label{GL}
{\cal L}_V W^{\cal M} =V^{\cal P}\partial_{\cal P} W^{\cal M} + \
(\partial^{\cal M}V_{\cal P}-\partial_{\cal P}V^{\cal M})W^{\cal P}\  .
\eea
The dilaton field $\varphi$ is incorporated through  density field 
$e^{-2d}=\sqrt{|g|} e^{-2\varphi}$
that transforms like a measure
\be
   {\cal L}_V e^{-2d}=\partial_P(V^{\cal P} e^{-2d})\, .
\ee
The algebra of generalized diffeomorphisms closes provided a set of
constraints is satisfied. 
The generalized diffeomorphisms allow to define the generalized dynamical  
fluxes\cite{reviewamn}
\beq \label{gfluxes}
    {\cal F}_{\cal ABC}= ({\cal L}_{E_{\cal A}}E_{\cal B})^{\cal M} E_{\cal 
CM}\, .
       \eeq
Fluxes  are totally antisymmetric in ${\cal ABC}$ (flat indices) and 
transform as 
scalars under generalized diffeomorphisms, up to the   closure constraints. 

In generalized Scherk-Schwarz compactifications 
\cite{effective,reviewamn,edft} the frame is split  into a space-time piece 
and an internal one. The former depends on the external $d$-dimensional 
coordinates\footnote{The  
${\tilde x}^\mu$ duals are dropped off, or equivalently the strong constraint 
is 
imposed in the space time sector.}  $x^\mu$ while the latter strictly depends on 
the internal $n$-dimensional (where $D=d+n$) 
coordinates ${\mathbb Y}^{I}=(\tilde 
Y_{i}, Y^{i})$, defined in the fundamental
representation of $O(n,n)$. Here ${I}=1, \cdots, 2n$ and
\beq \label{gss}
E_{A}(x, Y, \tilde Y)= {\cal U}_{A}{}^{A'}(x) E'_{A'}(Y, \tilde Y) \ .
\eeq
The matrix ${\cal U}$ encodes the field content  in the effective 
theory, while
$E'$ is a generalized frame that depends on the internal coordinates. All the 
dependence on the internal coordinates is through the frame. By using this 
splitting ansatz the generalized metric becomes $
H={\cal S}^{AB} {\cal U}_{A}{}^{A'} E'_{A'} {\cal U}_B{}^{B'} E'_{B'}={\cal
H}^{A'B'} E'_{A'} E'_{B'}$ where all the field dependence on space time 
coordinates is encoded in
\beq \label{Hmoduli}
{\cal H}^{A'B'}(x) ={\cal S}^{AB} {\cal U}_{A}{}^{A'}  {\cal U}_B{}^{B'} \ .
\eeq
 parametrizing the moduli space. In particular, we will deal with the 
 ``internal piece" ${\cal H}^{IJ}$, where  $I,J=1,...,2n$ are frame indices on 
the internal part of the double tangent space.

It proves useful to rotate to  a  
Right-Left basis ${\cal C}$ 
where  left and right coordinates are 
\bea
y_{Lm}&=&\frac12[(g+B)_{mn}y^n+\tilde y_m]\\\nn
y_{Rm}&=&\frac12[(g-B)_{mn}y^n-\tilde y_m]
\label{lrcoordinates}
\eea
in terms of $\mathbb Y_M=(y_m, \tilde y^m)$.
Namely, the rotation matrix  reads  
\be
R=  \left(\begin{matrix}
    (g+B)  & 1\\
    (g-B)& -1
  \end{matrix}\right) \ ,
  \ee
  and therefore 
from $E_A \rightarrow (E_{\cal 
C})_{A}=R_A{}^BE_B$   we see that 
$\eta$ becomes diagonal
\beq \label{etaC+C-}
(R \eta R^{T})_{AB}=\begin{pmatrix} 1_D & 0 \\ 0 & -1_D \end{pmatrix}\, .
\eeq
Since
the internal piece of $H$ lies in $O(n,n)/O(n)\times O(n)$ it is possible to 
show\cite{agimnr,Cagnacci:2017ulc} that the scalar matrix, in 
the Left-Right basis ${\cal C}$  can be written as  an expansion in scalar 
fluctuations
\beq \label{Hintc}
{\cal H}_{\cal C}= \begin{pmatrix}
    1_n+MM^T & M\\
  M^T & 1_n+M^TM
\end{pmatrix}+O(M^3)
\eeq
with $n^2$ independent degrees of freedom.

By using the expression for the generalized Lie derivative in the specific 
case of the frame
\be
{\cal L}_{E^{\prime}_A}E^{\prime}_B=\frac12 \big[E^{\prime\ 
P}_A\partial_PE^{\prime\ M}_B-E^{\prime\ P}_B\partial_PE^{\prime\ 
M}_A+\eta^{MN}\eta_{PQ}\partial_NE^{\prime\ P}_AE^{\prime\ Q}_B \big]D_{M}
\label{framederivative}
\ee
\beq
[E'_I,E'_J]={\cal L}_{E'_I}E'_J=f_{IJ}{}^{K} E'_{K} \ .
\label{structureconstants}
\eeq
 where the fluxes $f_{IJ}{}^{K}$, for the generalised Scherk-Schwarz reduction, 
must be 
constants and must  satisfy the constraints
\be \label{gsc}
f_{IJK}\equiv\eta_{KL} f_{IJ}{}^L=f_{[IJK]}\, ,\qquad f_{[IJ}{}^L
f_{K]L}{}^R=0\, .
\ee
The information  about the internal space is encoded in 
these constants. 
When replacing above results into the initial DFT action (\ref{dftaction}) the 
expression presented in (\ref{su2action}) is obtained.
\subsection{Enhanced gauge symmetry on the circle}
\label{sec:Enhanced gauge symmetry on the circle}

In Ref.\cite{agimnr} a  specific DFT frame was presented \footnote{The choice 
of frame was inspired by  a previous work \cite{dmr} set  in a different 
context. } 
 in order 
to reproduce the  effective action, obtained from string theory 
compactification on the circle, at 
the self-dual point.  As mentioned it requires to enhance the tangent space to  
$D=d+1+2$ but the frame only depends on the circle coordinate and 
its dual. In  a Cartan-Weyl basis the frame vectors read,
 \bea\nonumber
E_{\pm}&=& c( e^{\mp i\frac{2}{\sqrt{\ap}}y_L}  , ie^{\mp 
i\frac{2}{\sqrt{\ap}}y_L} 
, 0, 0, 0,0), \,\,\,E_{3}=  -c(0  ,0, 1  , 0, 0, 0,0)
\\
\bar E_{\hat \pm}&=& c(  0, 0, 0, e^{\mp 
i\frac{2}{\sqrt{\ap}} y_R}  , ie^{\mp 
i\frac{2}{\sqrt{\ap}}y_R} ,0) \,\,\,\, \,\bar E_{\hat 3}= 
-c( 0, 
0, 0, 0  ,0, 1 )
\label{sdframesu2}
\eea
 The directions $E_{\hat\pm}\equiv E_1+iE_2$ (and   $\bar E_{\hat \pm} = 
E_{\hat 1}+i E_{\hat 2}$ )  encode the 
extension of the tangent space.
 It is easy to check that, by using (\ref{framederivative}) (setting 
$c=i\sqrt{\ap}$) and by noticing that  
the only contributions to the partial derivative are
\be
\partial_A=(0,\ 0,\ \partial_{ y_L},\ 0,\ 0,\ \partial_{ y_R}),\
\ee
the $SU(2)_L\times SU(2)_R$ coupling constants (\ref{su2su2cc}) are 
obtained. In the Cartan-Weyl basis they read, 
\bea\label{sucoupconst}
\frac12f_{+-}{}^{3}&=& \frac12f_{\hat +\hat 
-}{}^{\hat 3}=f_{3+}{}^{+}=f_{\hat 3 \hat +}{}^{\hat +}=-f_{3-}{}^{-}=
-f_{\hat 3 \hat -}^{\,\,\,{\hat-}}=1
\\\nonumber
-\frac12 f_{\hat + \hat -}{}^{{3}}&=&-\frac12f_{+ -}{}^{\hat 3}=
f_{3\hat +}{}^{\hat +}=-f_{3\hat-}{}^{\hat -}=f_{\hat 3 
+}^{}{{+}}=-f_{\hat 3 -}{}^{{-}}=0
\eea
where we have used a hat to denote the indices constructed up from $4, 5,6$ 
Right indices. 

The construction of the frame is inspired in the coset structure 
(\ref{cosett1}) and on the structure of vertex operators in string 
theory\footnote{Basic ingredients and notation conventions for string theory 
vertices are briefly presented in Appendix.}.
Namely the correspondence among frame vectors and string current 
generators\cite{agimnr} can 
be established (here $e^{\pm L}=e_1\pm i e_2,\quad e^{\pm R}=e^1\pm i e^2$)
\bea
\bar { E}^\pm=e^{\mp \frac{2i}{\sqrtap}y^R} e^{\pm R} \ &\leftrightarrow&
\ e^{\mp \frac{2i}{\sqrtap}y^R(\bar{z})} d \bar z=\bar J^{\mp} d\bar z\ , 
\\
 \bar
{ E}^3=i/ \sqrtap\ d y ^R  & \leftrightarrow&  \  dy^R(\bar{z})= \bar J^{3} 
d\bar z \nn
\\
{ E}^\pm=e^{\mp \frac{2i}{\sqrtap}y^L} e^{\pm L} \ &\leftrightarrow& \
e^{\mp \frac{2i}{\sqrtap}y^L(z)}  dz =J^{\mp} dz \ ,\\
{E}^3=i/ 
\sqrtap\ d y ^L  
&\leftrightarrow& \  dy^L(z)=  J^{3} dz  \nn \ .
\eea
where $J^{\mp}(z)=e^{\mp \frac{2i}{\sqrtap}y^L(z)},  J^{3}(z)=\partial_zY(z)$ 
are the string currents satisfying the Operator Product 
Expansion (OPE) algebra of $SU(2)_L$ (and similarly 
for the Right sector). The corresponding string vertex operators $V^{i}(z, 
\bar z)$ for vectors  are
  \bea
V^{\pm,3}(z, \bar z)= i\frac{g'_{c}}{\alpha '^{1/2}}
\epsilon^{\pm,3}_{\mu}
  :  J^{\pm,3}(z)\bar\partial X^\mu e^{iK\cdot X}
  :\label{vertexvectors1}
 \eea
where $i=\pm,3$ and $K^\mu$ is the space time momentum. 

A similar construction was presented in \cite{waldramgaugings} (see also 
\cite{Blumenhagen:2014gva}) for the case of 
the $S^3$ reduction in the context of the  WZW model, inspired by 
\cite{schulz}. The purely  geometric case was studied in \cite{Lee:2014mla}. For 
the non-geometric one \cite{waldramgaugings},  the authors were able to show 
that allowing for a non trivial dependence on the dual coordinate of the Hopf 
fibre,  non-geometric gaugings  can be obtained \cite{Dall}. However, unlike 
the toroidal construction presented here (and in \cite{agimnr}) where  a 
clear world sheet picture arises,  the $S^3$  does not have 
non-contractible cycle and, therefore,  no winding states were really 
considered in  \cite{waldramgaugings}. 

For general compactification radios, the dependence on moduli is encoded in 
the exponential part of the vertex operators
\begin{eqnarray}
:  exp [i k_L y_L(z)+i k_R  y_R(\bar z)]e^{iK\cdot X}:
  \label{vt}
\end{eqnarray}
where 
\be
k_L^{(p,\tilde p)}=\frac{p}{R} +
\frac{\tilde p}{\tilde R}\, , \qquad
\ \  k_R^{(p,\tilde p)}=\frac{p}{R} -
\frac{\tilde p}{\tilde R}\, .
\ee
in terms of KK momenta $p$ and winding number $\tilde p$ satisfying the level 
matching condition 
$\bar N-N=p \tilde p$. $N= N_x+N_y$ ($\bar N= \bar N_x+\bar N_y$)
is the Left (Right) moving number operator, involving the sum of the number 
operator along the circle $N_y$ ($\bar N_y$) and the number operator for the 
non-compact space-time directions, denoted by $N_x$ ($\bar N_x$).
At the self-dual radio $R=\tilde R=\sqrt \ap$, the vertices separate into 
a Left part with  $k_R=0$ or into a Right vertices with $k_L=0$. The three 
vector states generating $SU(2)_L$ correspond to  $\bar N_x=1$, $N_x=\bar 
N_y=0$.
 The assignment $ N_y=1, p=\tilde p=0 $ corresponds to the KK (Cartan field) 
mode $A^3_{L\mu}$, while for $N_y=0, p=\tilde p=\pm 1$ (namely, $k_L^{(\pm 
1,\pm 1)}=\pm \frac
2{\sqrt{\alpha'}}; \, k_R^{(\pm 
1,\pm 1)}= 0 ) $  the charged vectors 
$A_{L\mu}^\pm$ are  obtained (and similarly for $SU(2)_R$).

When moving away from the fixed point, Left and Right parts mix up and, 
generically,  
the original vertex operator becomes ill defined as a conformal field. It must 
combine with other vertex operators, that have the same exponential 
contribution, in order to produce a new consistent 
vertex. Interestingly enough,  these combinations encode the Higgs mechanism by 
absorption of a 
vertex corresponding to a would be Goldstone boson field\cite{agimnr}.
With this picture in mind we generalize the frame (\ref{sdframesu2}) by 
including 
the dependence  
$k_L y_L+k_R  y_R  $ for the found values of momenta and windings.

\bea\label{genefra1}\nn
E_{\pm}&=& c( e^{\mp iw}  , \pm ie^{\mp iw} , 0, 0, 0,0)\qquad  \bar E_{+}= c(  
0, 
0, 0, e^{\mp i\bar w}  ,\pm ie^{i \mp \bar w} ,0) \\
E_{3}&=&  -c(0  ,0, 1  , 0, 0, 0,0)\qquad \qquad\,\, \, \bar 
E_{\hat 3}= -c( 0, 0, 0, 0 ,0, 1 
)\
\eea
where 
\bea
w&=& m_+y_L+m_-y_R,\qquad \bar w=m_-y_L+m_+y_R\label{ww}\label{genefra2}
\eea
and $
 m_{\pm} =k_R^{(1,\pm 1)}= \frac{1}{R}\pm \frac1{\tilde R}$.
Notice that $ m_+ \rightarrow \frac{2}{\sqrt{\ap}}$ at self-dual radio 
$R=\tilde 
R= {\sqrt{\ap}}$.
Again, by using (\ref{framederivative}) we obtain
\begin{eqnarray}\nn
 \big[E_+,E_- \big] &=&2\,({a_+}E_3 -{a_-}\bar E_{\hat 3}), \qquad \big[ 
E_{\hat +}, \bar E_{\hat -} 
\big]=2\,({a_+}\bar E_{\hat 3} -{a_-}E_3),\\\nn
  \big[E_3,E_+ \big]&=&a_+\,E_+, \qquad   \,\,\, \qquad \qquad 
\big[\bar E_{\hat 3}, \bar E_{\hat +} \big]=a_+\, \bar E_{\hat +} \\\nn
   \big[E_3,E_- \big]&=&-a_+\,E_-,  \qquad  \qquad\qquad \big[E_{\hat 
3}, \bar E_{\hat -} \big]=-a_+\, \bar E_{\hat -}\\\nn
  \big[\bar E_{\hat 3},E_+ \big]&=&{a_-}\,E_+, \,\,\ \qquad  \qquad \qquad  
\big[E_3, \bar E_{\hat +} \big]={a_-}\, E_{\hat +} \\
   \big[\bar E_{\hat 3},E_- \big]&=&-{a_-}\,E_-,  \qquad  \qquad\qquad \big[E_3,
\bar E_{\hat -} 
\big]=-{a_-}\, \bar E_{\hat -} 
\label{su2algebraaway}
    \end{eqnarray}
 with 
 \begin{equation}
  a_{\pm}=\frac{\sqrt{\ap}m_{\pm}}{2}.
  \label{amasmenos}
 \end{equation}
Thus, we find that,   by computing the fluxes 
(\ref{gsc}),  and up to a normalization factor ${\ap }^{\frac32}\sqrt2$, 
the constants proposed in (\ref{structureconstantsr}) are obtained (here 
written in a complex combination). Notice that, if  $R\rightarrow \tilde R$ 
then 
$a_-(a_+)\rightarrow 0(1)$ and the original $SU(2)\times SU(2)$  algebra is 
recovered. Moreover, it is easy to check that the algebra is invariant under 
T-duality transformation $R \leftrightarrow \tilde R$.

As mentioned, by systematically 
replacing the above structure constants (fluxes) into the general DFT action 
expression 
(\ref{dftaction}),  the exact 
spontaneously broken action, with $U(1)\times U(1)$ gauge symmetry, as computed 
from string theory (see Eq.(3.31) in \cite{agimnr}) is found.
In particular, vector fields $A_{L\mu}^{\pm}$ and  $A_{R\mu}^{\hat \pm}$ become 
massive, with 
masses $m_-$ by 
``eating'' the would be Goldstone bosons $\partial_\mu M_{\pm{\hat 3}}$ 
(and $\partial_\mu M_{3\hat \pm}$ ) that disappear from the spectrum. 

It appears instructive to see how some of the terms in the broken symmetry 
action arise.  For instance, by  inserting  the expansion in  
scalar fluctuations $M$ in the generalized scalar matrix \eqref{Hintc}, 
into the third row of the DFT action \eqref{dftaction} and using the values 
\eqref{structureconstantsr} for structure constants we find the quadratic the 
terms 
\begin{eqnarray}
&& 2 (m_+m_-+m_-^2)|M^{\pm\pm}|^2-  2 (m_+m_--m_-^2)|M^{\pm\mp}|^2\\\nn
 &=& \frac{4}{ 
R}m_{-}|M^{\pm\pm}|^2-\frac{4}{ 
R}m_{-}|M^{\pm\mp}|^2
\label{su2scalarmass}
\end{eqnarray}
reproducing the exact values $m^2_{\pm\pm}= \frac{4}{ 
R}m_{-}$ and  $m^2_{\pm\mp}=-\frac{4}{ 
R}m_{-}$ as computed from string mass formula 
\eqref{stringmasses}\footnote{Recall that, 
depending on the value of $R$ half of the scalars become tachyonic. This is an 
artefact associated to the ill defined bosonic string.}. 
The terms proportional to $m_+m_-$ and $m_-^2$  come from linear and quadratic 
terms in $M$ expansion in  \eqref{Hintc}, respectively.

In the same way it can be checked that the masses of the would be Goldstone 
bosons $M^{{\pm},\hat 3},M^{3\hat{\pm}}$ coincide, as it should, with the 
masses $m_-$ of the massive vector bosons.

Moreover, the same row in \eqref{dftaction} for  cubic terms in $M$ lead to 
\bea
&-&
  \frac{ 4}{{\sqrt{\ap}}}   M_{+-}M_{-+}M_{33}
({\frac{\sqrt{\ap}}{\tilde R})}^2
+
  \frac{ 4 }{\sqrt{\ap}}  M_{++}M_{--}M_{33}
{(\frac{\sqrt{\ap}}{ R})}^2
\\\nonumber
\eea
with $ m_+^2+m_-^2+2m_+m_-=(m_++m_-)^2=\frac4{{R}^2}$.

Coming back to the expressions \eqref{su2algebraaway}, it is worth noticing 
that the above brackets close  into a Lie  algebra for 
arbitrary values of $R$. Indeed,  by recalling that $f_{I J 
K}=\eta_{K L} f_{I J}^{\,\,\,\,\,\,\,L}$ are totally antisymmetric, it is easy 
to check that Jacobi identity is satisfied.
Of course, the found algebra should correspond to one of the known semi-simple 
algebras. Since it involves six charged generators and two Cartan 
ones the only possibility is $SU(2)\times SU(2)$. Actually, this can be 
explicitly shown by performing the  linear combinations 
of generators
\begin{equation}
\begin{aligned}
E'_{\pm} =& E_{\pm}; \qquad \bar E'_{\hat \pm} = \bar E_{\hat \pm} \\
E'_{3} =& a_{+} E_{3} - a_{-} \bar E_{\hat 3}\\
E'_{\hat 3} =& -a_{-} E_{3} + a_{+} \bar E_{\hat 3} ,
\end{aligned}
\end{equation}
namely a rotation by the $O(3,3)$ matrix
\begin{equation}
\begin{pmatrix}
1_2 & 0 & 0 &  0\\
0  & a_{+} & 0 &  -a_{-}\\
0  & 0 & 1_2 &  0\\
0 & -a_{-} & 0  & a_{+}
\end{pmatrix}
\end{equation}
and using that $a^{2}_{+} 
- a^{2}_{-}=1$.

We thus see that, even in the broken phase, there 
is still an  underlying $SU(2)$ symmetry (now mixing massive and massless 
states). However, once the above 
frame is chosen, the $O(3,3)$ full symmetry gets broken and, therefore, it can  
not be rotated to the starting point.
Recall also that, in terms  of fields, the combination of $U(1)$ gauge bosons 
\beq
A^{3'{\mu}} =a_{-} A_{L}^{3\mu} + a_{+} A_{R}^{\hat 3\mu}=  V^{\mu}+ 
B^{\mu}\\
\label{Aprime}
\eeq
is the right combination in terms of
\begin{equation}
V_{\mu}=\frac1{2R}({A_{\mu}^{3}
 +\bar A_{\mu}^{3}}) \ , \qquad
   B_{\mu}=\frac1{2\tilde R}({A_{\mu}^{3}
 -\bar A_{\mu}^{3}})
 \end{equation}
 which are the KK reductions of the metric and antisymmetric 
fields and  with respect to which massive states carry 
integer charge (see \cite{agimnr}).

It is instructive to look at the above results from the string theory point of 
view. There, the structure constants can be essentially read from the 3-gauge 
vector bosons vertices with vertex operators $V^i $.
For the massless case they read (see  \cite{agimnr} 
for notations and explicit computations), 
 for Left vectors,  
\begin{eqnarray}\nn
  <V_{L}^i V_{L}^j V_{L}^k>&=& \pi
g_{c}\frac{i}{\sqrt{\ap}}\epsilon^{ijk}\left[(\epsilon^{k}_{3}\cdot
K_{1})(\epsilon^{i}_{1}\cdot\epsilon^{j}_{2})-(\epsilon^{j}_{2}\cdot K_{1})
    (\epsilon^{i}_{1}\cdot\epsilon^{k}_{3})\right.
    \\\nn
    &+&\left.(\epsilon^{i}_{1}\cdot 
K_{2})(\epsilon^{k}_{3}\cdot\epsilon^{j}_{2})\right]\nn
\end{eqnarray}
where $K_{1},K_{2},K_{3}$ are the space time momenta of vertices $i, j, k$ 
respectively. 
Namely, we can read the 
$\epsilon^{ijk}$ structure constants of $SU(2)_L$ (and 
similarly for $SU(2)_R$) and there is no mixing between L-R sectors.

On the other hand, away from the self-dual point we find the three-point 
coupling of Left and Right vectors can be written as  
\begin{eqnarray}\nn
\langle V^{+}_{L}V^{-}_{L}V^{3}_{L}\rangle &=&
\frac{\pi g'_{c}}{2\sqrt{\ap}}(a_+)E(k_i,\epsilon_i) \\\nn
\langle V^{+}_{L}V^{-}_{L}V^{3}_{R}\rangle &=&\frac{\pi 
g'_{c}}{2\sqrt{\ap}}(a_{-})E(k_i,\epsilon_i)\\\nn
\langle V_R^{+}V_R^{-}V^{3}_{L}\rangle &=& \frac{\pi 
g'_{c}}{2\sqrt{\ap}}(a_{-})E(k_i,\epsilon_i)
\label{3psu2}
\end{eqnarray}
where $E(K_i,\epsilon_i)=
(\epsilon^{'}_{1+}\cdot\epsilon^{'}_{2-})(K_{1}
\cdot\epsilon^{}_{3}) + (\epsilon^{'}_{1+}\cdot\epsilon^{}_{3})(K_{3}
\cdot\epsilon^{'}_{2}) + (\epsilon^{}_{3}\cdot\epsilon^{'}_{2-})(K_{2}
\cdot\epsilon^{'}_{1+}) $ is a factor that depends on space time momenta and 
vector polarizations.
Thus, if by analogy with the  dual point case,  we interpret the coefficients  
as 
the moduli dependent coupling constants we find; $f_{+-3}(R)=a_+, f_{+-\hat 
3}(R)=a_-$ 
etc.
Moreover,  by considering the 
combinations ($\ref{Aprime}$) above, we can again identify  the underlying 
$SU(2)$ 
structure.
The $SU(2)$ controls the allowed three point functions through conservation of 
internal Right and Left momenta.

\section{Enhancing-breaking of gauge symmetries for \\generic 
toroidal compactifications}
\label{sec:extensions}
In this section we briefly discuss possible realizations of the enhanced 
symmetry breaking mechanism, through moduli dependent structure constants,  for 
general toroidal compactifications. Bosonic string compactification 
\cite{Giveon:1994fu}  on a 
$T^r$ torus  of  $r$ 
dimensions gives  rise to a gauge symmetry group $G_L\times G_R$  of 
rank $2r$ ($r$ coming from Left and Right vectors associated 
to the metric and $B$ field degrees of freedom). At generic points of the 
compactified manifold this group is simply $U(1)^r_L\times U(1)^r_R$ but, at 
special moduli points,  $G_L$ is a non abelian group with $dimG_L=n=n_c+r$. 
Here 
$n_c$ counts the number of charged generators associated to the presence of 
non trivial winding and KK momenta.
By reasoning as in the circle case, if we assume that the number of 
massless degrees of freedom at some point of enhancing  is 
given by 
\begin{equation}
 dim( \frac{O(d+n,d+n)}{O(d+n)\times O(d+n)})=d^2+2nd +n^2 
 \label{oddstructure}
\end{equation}
this appears to  correspond to the $d^ 2$ degrees of freedom of gravity 
(plus 
B field),  the $2n$ vectors of a $G_L\times G_R$  and $n^2$ 
scalars in bi-adjoint representations.
If $n=r$ it gives the correct counting for $U(1)_L^r\times U(1)_R^r$ degrees of 
freedom.

For a circle compactification $r=1$ and  by choosing  $n=2+1$ the counting 
corresponds to an $SU(2)_L\times SU(2)_R$ gauge group with  scalars in the 
$(\bf 3,\bf 3)$ representation,  as it is the 
case for the self-dual point $R=\tilde R$.

For a $T^2$ toroidal compactification, when  $n=2\times 3$, the number of 
massless degrees of freedom for 
$ SU(2)_L\times SU(2)_L\times SU(2)_R\times SU(2)_R$ with scalars in 
$(\bf 3,1,\bf 3,1)+(1,\bf 3,1,\bf 3)+(\bf 3,1,1,\bf 3)+(1,\bf 3,\bf 3,1)$,  
corresponding to a possible torus enhancing point,  is reproduced. Also, the 
correct counting occurs for $n_c=6$,   for the 
degrees of freedom  of $SU(3)_L\times SU(3)_R$ with scalars in the $(\bf 8,\bf 
8)$ representation at the point of maximal enhancing \cite{Giveon:1994fu}.

The generalization  of the exponential contribution (\ref{vt}) to the string 
vertex 
operators for a general torus (with lattice vectors $e^a{}_m$) reads (see 
Appendix for notation)
\begin{eqnarray}
:   e^{i k_L. y_L(z)+i  k_R .y_R(\bar z)}\, e^{iK\cdot X}:
  \label{vtg}
\end{eqnarray}
with  Left and Right momenta
 \bea
k_{L}^{a} &=&e^a{}_m p_{L}^{m}, \qquad k_{R}^{a} =e^a{}_m p_{R}^{m}\nn
\label{klkr}
\eea
where 
\be
p^m_L=\tilde p^m
+g^{mn}(p_n-B_{nk}\tilde p^k)\, ,\qquad
p^m_R= -\tilde p^m
+g^{mn}(p_n-B_{nk}\tilde p^k)\, .
\label{plpr}
\ee
 $g_{mn}=e^a{}_me^a{}_n $ defines the internal metric whereas $B_{mn}$ are 
the  internal components of the Kalb-Ramond field.

Notice that, by using \eqref{lrcoordinates} the following relation holds
 \begin{equation}
 k_L. y_L+  k_R .y_R=\mathbb P.\mathbb Y
 \end{equation}

Gauge symmetry enhancing occurs at specific values of moduli 
$(g_0,B_0)$, encoded in the frame vectors 
$e_m^a(g,B)$  and of windings and momenta (encoded in the generalized 
momentum $\mathbb P=(p_1,p_2,\dots;\tilde p^1,\tilde p^2\dots$). At such 
values,   $k_{L}^{a}$  become roots of a semi 
simple 
algebra ($k_{R}^{a}=0$) and similarly for the right sector.
Namely, at such points, the internal part of vertex operators in (\ref{vtg}) 
becomes
\begin{equation}
 E_{\alpha}\simeq e^{i k_L^{\mathbb P}. y(z)}
 \label{eq:chgen1}
\end{equation}
 with   $k_L^{(\mathbb P)\, a}\equiv 
\alpha_m^a$ a root of the semi simple algebra and where  $m=1,2,\dots$ 
(associated to  ${\mathbb P}$ values) labels the  charged 
operators.  These vertex operators, together 
with the corresponding Cartan operators, close the  OPE of a $G_L$ group affine 
algebra.

Let us consider the 2-torus example discussed in the Appendix.  
For generic values of $E=g+B$  the gauge group is $U(1)_L^2\times 
U(1)_R^2$ but  enhancings occur  at different points 
\cite{{Giveon:1994fu}}. 
For 
instance, by choosing the basis\footnote{The $\sqrt2$ is just a normalization 
factor in order to keep the usual convention for $\alpha^2=2$ for the roots.} 
$e_m  =\frac{1}{\sqrt2}\alpha_m$ with $m=1,2$ with  $\alpha_{1,2}$ the simple 
roots of $SU(3)$ and 
$B_{12}=g_{12}=-\frac12$ we see that there are six generalized momentum 
vectors
\begin{eqnarray}
 \mathbb{P}=
\pm(1,0,1,0),\pm (-1, 1,0,1),\pm (0, 1,1,1)
\label{Psu3}
\end{eqnarray}
that satisfy 
 the LMC,  and such that  $P_R^m=0$. They give rise to six 
extra massless states with 
\be
 P_L=\pm(1,0),\pm(0,1), \pm(1,1)
\ee
 Similarly   $\mathbb{P}=
\pm(-1,1,1,0),\pm (0, -1,0,1),\pm (1,0,-1,-1)$  lead to same roots for $P_R$  
while $P_L=0$. At the end, the   enhanced 
$SU(3)_L\times SU(3)_R$ gauge 
group is generated.

Also, for 
\begin{equation}
G_{m n}=\begin{pmatrix}
1 && 0\\
0 && 1
\end{pmatrix}
\end{equation} and $B=0$ an enhancing to $ (SU(2)\times SU(2))_L\times 
(SU(2)\times SU(2))_R$ is obtained for $\mathbb{P}=(\pm 1,0,\pm 1,0),(0,\pm 
1,0,\pm 1) $, etc.

The description of the enhancing-breaking of the gauge symmetry could be, in 
principle, described by generalizing the steps  presented in 
the previous section for the circle particular situation. We have not pursued 
this construction systematically but 
we present some examples for the 2-torus case\footnote{A systematic 
derivation is proposed in \cite{Cagnacci:2017ulc} with a modification of 
the generalized Lie derivative.}.
For a general  $r$-torus compactification, from a DFT point of view, we 
should 
consider a doubling of the internal manifold $T^r\times \tilde T^r$,  
incorporating  both tori coordinates 
$y^m$  as well as their duals
$\tilde y_m$ with $m=1,\dots r$ ( in a $O(r,r)$ writing it corresponds to the 
double coordinate  $\mathbb Y^M $ with 
$M=1,\dots 2r$.).

Following the counting (\ref{oddstructure}) it appears that the tangent frame 
must be  enlarged further in order to incorporate information about charged 
operators.  Thus, if we were to describe a $G_L\times G_R$  point, 
besides the $r+r$ frame vectors, associated to  the internal 
coordinates Cartan generators, $2n_c$  extra frame vectors should be 
incorporated with $n_c= (dimG-r)$  associated to the left charged generator 
vertices 
\eqref{eq:chgen1} (and similarly for  Right vertices). Thus, in principle, 
we 
should have a $2 dimG$  tangent frame space  where the 
frame vectors only depend on $\mathbb Y$ internal coordinates. Each frame 
vector could be written in a given $2dim G$ basis and frame vectors 
associated to charged operators are expected to depend on an exponential factor 
 $e^{i k_L^{(\mathbb P)}. y_L}$ (and similarly for  R vectors) where $\mathbb 
P$ encodes the specific values of momenta and winding characterizing the 
enhanced vectors.
 
 For generic points in the compact manifold we will have internal directions 
$e_m^a(g,B)$ depending on the moduli fields and, therefore,  so do  
$k_L^{(\mathbb 
P)}(g,B)$ and $k_R^{(\mathbb 
P)}(g,B)$\footnote{We avoid writing the dependence on moduli in order to 
lighten the notation}. At selected values of $g, B$ these 
directions become the simple roots of the enhancing algebra.
Therefore, away from fixed points we expect the frame vectors to depend on both 
$e^{i k_L^{(\mathbb P)}. y_L+i k_R^{(\mathbb P)}. y_R}$,  as in fact, we 
found in the circle case (recall that the possible values of $\mathbb P$ 
are fixed). When moving into the 
 fixed point,  $\mathbb P$ values will produce the roots of $G_L$ (and 
$k_R^{\mathbb P}=0$) and the roots of $G_R$ (and 
$k_L^{\mathbb P}=0$
).
This is indeed what we found in the circle case and we now illustrate in its 
simplest generalization of the $2-$torus case near the $SU(2)^4$ fixed point.

Let us name  $\mathbb Y^M=(\tilde y_m, y^m)$, $m=1,2$,  the double torus 
coordinates or $( y_{Lm},y_{Rm})$  in a $L-R$ basis. The exponential 
contributions can now be written in terms of  
$e^{i \theta_j}$ where 
\begin{equation}
\theta_{(j)} =k^{m}_{(j) L} y_{Lm} +k^{m}_{(j) R} y_{Rm} =k^{1}_{(j) L} y_{L1} 
+ k^{2}_{(j) L} y_{L2} + k^{1}_{(j) R}  
y_{R1} + 
k^{2}_{(j) L}  y_{R2}
\end{equation} 
Here $(j)$ encodes the $\mathbb P=(p_1,p_2, \tilde p^1, \tilde p^2)$ values 
that would lead to  $SU(2)_j$ at the self-dual point.
For instance,   $\mathbb P=(\pm 1,0,\pm 1,0)$ generates a $k^{m}_{(1)L}$ 
 and $k^{m}_{(1)R}$ 
(where  $k^{m}_{(1)R}=0$ at self-dual point) etc.
Overall we find 
\bea
\mathbb P&=&(\pm 1,0,\pm 1,0)\rightarrow  k^{m}_{(1)L(R)}\\\nn
\mathbb P&=&(0,\pm 1,0,\pm 1)\rightarrow  k^{m}_{(2)L(R)}\\\nn
\mathbb P&=&(\pm 1,0,\mp 1,0)\rightarrow  k^{m}_{(3)L(R)}\\\nn
\mathbb P&=&(0,\pm 1,0,\mp 1)\rightarrow  k^{m}_{(4)L(R)}
\label{psu4}
\eea
where at the corresponding self-dual point $ k^{m}_{(1)R}= k^{m}_{(2)R}=0$ and 
$ k^{m}_{(3)L}= k^{m}_{(4)L}=0$.
Following the general steps sketched above we thus propose a 
generalized twelve dimensional ( $2dimG_L=12$) frame with frame 
vectors depending only on $\mathbb Y^M$. A 
straightforward generalization of the circle case leads us to the frame vectors

\bea
E_{+(j)}&=&\big(0^{3(j-1)}; \vec{v}_{+(j)}; 0^{3(4-j)}
\big)e^{-i\theta_{j}}=E_{-(j)}^*\\
E_{0(j)}&=&\big(0^{3(j-1)}; \vec{v}_{0(j)}; 0^{3(4-j)})
\label{framesu4}
  \eea 
  where $\vec{v}_{\pm j}=(0,1,\pm i)$ ($\vec{v}_{0j}=(i,0,0) $) is a 3 dim 
vector inserted at position $j$.  Notice that $E_{+(j+3)}\equiv \bar E_{+(j)}$ 
correspond to Right vectors. At  the self-dual point these vectors lead to 
$SU(2)_L\times SU(2)_R$ algebra for each value of $j$.

Moving away from the $SU(2)^4$ fixed point generically mix the twelve 
generators leading to moduli dependent structure constants $f_{IJK}(g,B)$ ($I,J,K=1,\dots 12$). Actually, due to the frame structure (\ref{framesu4}), the 
mixing occurs between Left and Right components for a given value of $(j)$, 
namely for the same would be $SU(2)_j$ frame.

For instance, by setting for simplicity for $B=0$  but for generic metric, we 
find
\bea
f_{+-\bar 0}(1)(G) &\propto & k_R(1) = \sqrt{2}[G_{11} + \frac{G_{22}}{det(G)} 
- 2]^{\frac12}\\
f_{+-\bar 0}(2)(G) &\propto& k_R(2) =\sqrt{2}[G_{11} + \frac{G_{11}}{det(G)} 
- 2]^{\frac12}
\eea
which generalizes the expression (\ref{structureconstantsr}) found for the 
circle.
By inserting these constants into the generic DFT action it is possible to 
check, as 
sketched in the introduction,  that the action for a generic spontaneous 
symmetry breaking to $U(1)^4$ is achieved. The complete computation was 
performed 
by using a computer program.

 The masses of the Left-vectors bosons are 
 \begin{eqnarray}
  m^{2}_{1}&=& f^2_{+-0}(1)(G)\\
  m^{2}_{2}&= & f^2_{+-0}(2)(G)
 \end{eqnarray}
 and (similarly  for the R-vectors). They  coincide   with the masses computed 
from string theory (\ref{stringmasses}).
  The values  $G_{12}=0, G_{11}=G_{22}=1$ lead to   $m^{2}_{1}=  m^{2}_{2}=0$ 
thus leading to the $SU(2)^4$ enhancing. Also, $G_{12}=0,G_{11}=1, 
G_{22}=(\frac{R_{(2)}}{\sqrt{\ap}})^2$ corresponds to a partial breaking stage 
to $SU(2)_{1L 
}   \times U(1)_{2L} \times SU(2)_R  \times U(1)_{2R}$ etc.

Recall that, generically, for a given point of enhancing $(g_0,B_0)$ with 
$G_L\times G_R$ gauge group,  once   the values of fluxes $f_{ABC}(g,B)$ are 
found, we just have to plug them into the  DFT action to obtain the effective 
gauge symmetry  broken action.
We have shown how to compute these fluxes from a generalized tangent frame 
construction. However, we can easily read them  from string theory 3-vector 
bosons amplitudes, as we saw for the circle case.
Namely, at a given fixed point, as mentioned $k_L^{(\mathbb 
P)}(g_0,B_0)=\alpha^{(\mathbb 
P)}(g_0,B_0)$ become simple roots of the $G_L$ group algebra and 
$K_{L(R)}^{(0)}=0$ 
for Cartan vectors.
Let us consider  the 3-point amplitudes for massless bosons. For charged bosons 
we can write, up to an antisymmetric factor in vertex indices depending on 
vector polarizations (see (\ref{3psu2})), as 
 \begin{eqnarray}
\langle V({k_L^{(\mathbb P_1)}})
V({k_L^{(\mathbb P_2)}})
V({k_{L}^{(\mathbb P_3)}})\rangle
 &\propto & f_{ \alpha^{(\mathbb 
P_2)}\alpha^{(\mathbb 
P_2)}\alpha^{(\mathbb 
P_3)}}(g,B)
\end{eqnarray}
where $f_{ \alpha^{(\mathbb 
P_2)}\alpha^{(\mathbb 
P_2)}\alpha^{(\mathbb 
P_3)}}(g,B)=1$ if $\mathbb P_3=-\mathbb P_1-\mathbb P_2$ and vanishing  
otherwise 
due to momentum conservation. The constants are antisymmetric. At the self-dual 
point this indicates  that structure constants $f_{ 
\alpha_1\alpha_2\alpha_3}$ vanish unless  $\alpha_1+\alpha_2$ is a root 
(and similarly for Right the  sector). 
For the same reason mixings of Left and Right indices vanish.
On the other hand, by denoting by $V(I_{L(R)})$ with  $I=1,\dots r$ the Cartan 
vectors,    the only non vanishing amplitudes are  
\begin{eqnarray}\nn
\langle V({k_L^{(\mathbb P)}})
V({k_L^{(-\mathbb P)}})
V(I_{L(R)})\rangle
&\propto & k_{L(R)}^{(\mathbb 
P)}(g,B)_I 
\label{structureconstvertices}
\end{eqnarray}
and, by 
identifying the amplitudes coefficients with algebra structure constants, we 
have
\begin{eqnarray}
 k_{L}^{(\mathbb 
P)}(g,B)_{I_L}&=& f_{ \alpha^{(\mathbb 
P)}\alpha^{(\mathbb 
-P)}I}(g,B),\qquad
k_{R}^{(\mathbb P)}(g,B)_{I_R} = f_{ \hat \alpha^{(\mathbb 
P)}\hat \alpha^{(\mathbb 
-P)}\hat I}(g,B)\\\nn
k_{L}^{(\mathbb 
P)}(g,B)_{I_R}&=& f_{ \alpha^{(\mathbb 
P)}\alpha^{(\mathbb 
-P)}\hat I}(g,B),\qquad
k_{R}^{(\mathbb P)}(g,B)_{I_L} = f_{ \hat \alpha^{(\mathbb 
P)}\hat \alpha^{(\mathbb 
-P)}I}(g,B)\end{eqnarray}
where we have used hatted  indices for Right generators.
Thus, we propose the algebra
\bea\nn
 \big[ E_{\alpha},E_{-\alpha } \big] = k_{L}^{(\alpha)I} 
H_I+k_{R}^{(\alpha)\hat I}\hat H_{\hat I} 
&\qquad&     \big[ \hat E_{\hat \alpha}, \hat 
E_{-\hat \alpha } \big] = k_{L}^{(\hat 
\alpha)I} H_{I}+k_{R}^{(\hat \alpha)I} \hat H_{I}             \\\nn
\big[ H_{I},E_{\alpha } \big] = k_{L}^{(\alpha)I} E_{\alpha}&\qquad& 
\big[ \hat H_{\hat I}, \hat E_{\hat \alpha } \big] = k_{R}^{(\hat \alpha)\hat 
I} \hat 
E_{\hat \alpha} \\
\big[ H_{I},\hat  E_{\hat \alpha } \big] = k_{L}^{(\hat \alpha)I} \hat 
E_{\hat 
\alpha}&\qquad&
\big[ \hat H_{ I},E_{\alpha } \big] = k_{R}^{(\alpha)I} 
E_{\alpha}
\label{generalalgebra}
\eea
where we have used $\alpha=\alpha^{(\mathbb 
P)}$  to alleviate the notation.
It is easy to show that \eqref{generalalgebra} satisfies  Jacobi identities.
and therefore defines a Lie algebra.

At the self dual point (where 
$k_{R}^{\alpha}(g_0,B_0)=k_{L}^{\hat \alpha}(g_0,B_0)=0 $) and $f_{ 
\alpha -\alpha I}=  \alpha 
^I$, (and similarly for Right sector) the algebra reduces to  to the 
gauge algebra of $G_L\times G_R$ in the Cartan-Weyl basis.
For instance notice that  $[E_{\alpha},E_{-\alpha}]=\alpha^ 
IH_I$ for charged generators $E_{\alpha}$ and Cartan generators $H_I$, as 
expected.

As an example let us specify to the  $SU(3)_L\times SU(3)_R$ case (the 
expressions 
are, however, general).
  Since this algebra must be continuously connected with the $SU(3)_L\times 
SU(3)_R$ algebra at the fixed point and has four  Cartan generators the only 
possibility left is an  $SU(3)\times SU(3)$. Again, away from the fixed point,  
we detect the same  underlying algebra, now mixing massive and massless 
(associated to Cartan generators) vector fields.

Let us underscore that, by replacing above fluxes into the DFT action 
\eqref{dftaction} and by performing the scalars  expansion 
\eqref{Hintc}, as we did for the circle case example, the full broken  
$G_L\times G_R$ symmetry action is found. Recall that this is valid for an 
arbitrary fixed point in a general $r$ dimensional toroidal compactification.
As a check we show in the  Appendix that the resulting masses for 
vector fields and scalar fields, as functions of moduli, coincide with the 
string theory ones.

\section{Summary and outlook}
\label{sec:Summary and outlook}
A well known distinguished feature of string 
theory is the enhancing of gauge symmetries at certain values  
of moduli backgrounds.
In this work we have shown that DFT formulation helps to identify 
an interesting description of enhancing phenomena. Namely, enhancing 
information appears encoded into moduli dependent generalized fluxes 
$f_{ABC}(g,B)$ with $A,B,C=1,\dots 2n$ indices in an $O(n,n)$ vector 
representation. Splitting indices in a Left-Right basis $A=(a,\hat a)$,  
it appears that enhancing occurs for moduli values $(g_0,B_0)$ such that  
generalized fluxes with mixed indices vanish. In this situation 
$f_{abc}(g_0,B_0)(f_{\hat a\hat b \hat c}(g_0,B_0))$ become the structure 
constants of a $G_L (G_R)$,  $dim G_L=n$ dimensional non-Abelian gauge group.  
In fact,  the vector boson masses are proportional to mixed indices 
fluxes (\ref{massvb}).

As mentioned, when replacing these moduli dependent fluxes into the generic DFT 
action the effective string theory action is reproduced, as long as  up 
to  slightly massive states are kept.  
Therefore,  DFT is providing us with a generic field theory action  
that leads to an accurate description of string theory results even in a non 
trivial stringy situation  of gauge symmetry enhancing-breaking when  massive 
states with associated momenta and winding are present. 
As discussed in \cite{agimnr} for the circle case (and extended in 
\cite{Cagnacci:2017ulc} for other situations) by giving $vev$'s to some 
specific scalar fields, the string broken symmetry action 
can be approximately  obtained  as an expansion in powers of the $vev$,s.   
It is worth insisting that the DFT construction we are presenting 
here is already producing the broken symmetry phase. Moreover, different 
coefficients and masses in the string action are exactly reproduced as 
functions of moduli and  not as an expansion.

In addition,  we have shown (at least for some examples) that generalized 
fluxes 
can be computed by introducing a generalized frame in tangent space with 
extended tangent directions but depending only on the coordinates  of the 
double ``physical torus''. The DFT generalized Lie 
algebra closes even though the strong constraint is not satisfied. In fact, the 
frame is explicitly non-geometric since it is a  function of the double 
coordinates $\mathbb{Y}=(Y,\tilde Y)$.

The idea of doubling the number of coordinates in order to describe winding 
modes was one of the original motivations of DFT. However, only recently 
windings were actually included in DFT.  In \cite{agimnr} a step in this 
direction was performed by showing that DFT can describe the massless sector of 
an enhanced gauge symmetry situation with windings playing  a fundamental 
role and where  an unpaired number of Left and Right $N-\bar N=\pm 1$ moving 
oscillators is implied in string theory (see also \cite{Cagnacci:2017ulc}). 
Also in \cite{Aldazabal:2016yih} a generalized  
KK toroidal compactification (GKK) of DFT containing towers of massive states  with 
generic  windings and KK momenta was considered, for the case  $N-\bar N=0$, 
namely with the level matching condition $\mathbb P ^2= 0$.   
The present work is a contribution in between,  in the sense that it incorporates 
slightly massive states with paired and unpaired oscillators but disregards 
higher massive states. 

The tangent space extra dimensions in the above construction  are associated to 
states with non vanishing momenta and windings,  actually with $\mathbb P^{ 
2}=\pm 1$. It may appear somewhat awkward that moving continuously 
from one 
point of enhancing to another  could lead to a discrete change in the 
number of these extra tangent dimensions, even if these are just 
tangent directions and not physical dimensions at all.
In string theory the vector fields that become massless to lead to gauge 
enhancing are part of the spectrum and they are associated to $N-\bar N=\pm 
1$. It appears that in this situation DFT in lower dimensions should 
allow for the presence  of new vector fields, say   
$A_{L(R)}^\nu(x,{\mathbb Y})$ where ${\mathbb Y}$ are coordinates on a double 
torus.

A possible way  these jumps could be actually understood is through a GKK mode 
expansion, as considered in \cite{Aldazabal:2016yih}, but 
allowing for states with LMC  $\delta (\mathbb P^{ 2})=\pm 1,0$. 
For instance, 
\bea
A_{L\nu}(x, {\mathbb Y})
&=&\sum_{\mathbb P}
A_{L{\nu}}^{({\mathbb P})}(x) e^{i \mathbb{P}_{M} \mathbb{Y}^{ M}}\,\delta 
(\mathbb P^{ 2},1)\\\nn
&=&  \sum_{\mathbb P}
A_{L\nu}^{I({\mathbb P})}(x) e^{i k_L.y_L+k_R.y_R}\,\delta 
(\mathbb P^{ 2},1),\\
\eea
where $P_L, P_R$ depend on moduli (\ref{plpr}). When moving continuously along 
the moduli space, for certain  values of $ \mathbb P$, GKK modes $k_R=0$ and 
the 
corresponding fields $A_{L{\nu}}^{({\mathbb P})}(x)$ become massless. For 
instance for the $T^2\times \tilde T^2$ the six modes (\ref{Psu3}) become 
massless for $g_{11}=g_{22}=-2 g_{12}=-2B_{12}=1$ leading to the charged 
operators of $SU(3)_L$. Sliding  away 
from this point  the
masses of these modes vary continuously from zero. When reaching 
the moduli point $g_{11}=g_{22}=1 ;B_{12}=0$ other modes (the six 
modes shown in (\ref{psu4}))  become massless\footnote{Notice that 
there are two common modes $\mathbb P=(\pm 1,0,\pm 
1,0)$.} and lead to $SU(2)_L ^2$ enhancing.
The massless vector fields are those  captured 
by the extended tangent frame vector in DFT. 
Moreover, we saw that at  the neighbourhood of the point of enhancing 
associated to a  gauge generator algebra  $G$,  there is still an underlying  
global $G$ algebra, mixing massless (Cartans) and slightly massive 
states. When moving away from that point other fields, now with comparable 
masses,  will come into play and will have non neglectable 3-point amplitudes 
indicating a possible infinite enhancing of the global algebra.
This appears to be an indication of the presence of a Generalized 
Kac-Moody algebra of the kind discussed in \cite{Aldazabal:2016yih} but 
including unpaired LMC conditions. Of course these ideas need further 
investigation.

For the sake of simplicity we have dealt with the bosonic string example. 
However the reasoning should be straightforwardly applicable to  the (bosonic 
sector) of  Heterotic theories (\cite{hk}) or Type II theories obtained from 
U-dual Extended Field Theories (EFT)\cite{eft}.
It 
could also be interesting to explore the inclusion of extra tangent 
dimensions directly in gauged supergravity 
theories \cite{Samtleben:2008pe,Trigiante:2016mnt}.

\section*{Acknowledgments}
We thank   G. 
Torroba, F. Schaposnik Massolo 
and  D. Marqu\'es for useful discussions and 
comments and, in particular, C. Nu\~nez who participated in the first steps of 
this research.
This work was  supported
by PIP CONICET grant,  PICT-2012-513 and by IBS-R018-D2.
G. A. thanks the Instituto de F\'isica Te\'orica (IFT
UAM-CSIC) in Madrid for its support via the Centro de Excelencia Severo Ochoa
Program under Grant SEV-2012-0249 and  A.S.ICTP
for  hospitality and partial support during the first steps  of this 
work.
\bigskip

\noindent
{\bf Note added in proof:}
The same day this paper was made public, the article 
\cite{Cagnacci:2017ulc} appeared adressing similar issues from a rather 
complementary point of view.

\appendix
\section{Vertex operators and enhancing}
\label{sec:Vertex operators and enhancing}

We summarize here some string theory ingredients needed in the body of the 
article.

A generic vertex operator contains an exponential contribution that can be 
written in terms of Left and Right moving coordinates
$y_L(z),y_R(\bar 
z)$as  $e^{ik\cdot X+ik_L\cdot y_L+ik_R\cdot\bar y_R}:$ where  
$K^{\mu}$ stands for the space-time momentum while $k_{L(R)}$ are the internal 
$L(R)$ momenta.
It is convenient to use coordinates $y^a_{L(R)}=e_m{}^a y^m_{L(R)}$ with
tangent space indices $a, b, ...$, defined in terms of the vielbein
$e_m{}^a$ ($\delta^{ab}=e_m{}^a g^{mn}e_n{}^b$) since they
have the standard OPEs.
Namely, the propagators read
\bea
\langle X^\mu(z,\bar z) X^\nu(w,\bar w)\rangle &=& -
\frac{\alpha '}2\eta^{\mu\nu}ln|z-w|^2\, ,\nn \\
\langle Y^a (z) Y^b(w)\rangle= -\delta^{ab}
\frac{\alpha '}2 ln(z-w)\, ,&&\,
\langle \bar Y^a(\bar z) \bar Y^b(\bar w)\rangle = -\delta^{ab}
\frac{\alpha '}2 ln(\bar z-\bar w)\, .\nn
\eea
and the vertex operator momenta
are
\bea
k^a_{L}=e^a{}_m p^{m}_L\, ,\qquad
k^a_{R}=e^a{}_m p^{m}_R ,
\eea
where
\be
p^m_L=\tilde p^m
+g^{mn}(p_n-B_{nk}\tilde p^k)\, ,\qquad
p^m_R= -\tilde p^m
+g^{mn}(p_n-B_{nk}\tilde p^k)\, .\nn
\ee
 The  stress energy tensor is
 	\bea
 T(z)  &=&-\frac 1{\alpha '}(\eta_{\mu\nu}:\partial_z 
X^\mu(z)
 \partial_z X^\nu(z):+\delta_{ab}:\partial_z Y^a(z)\partial_z Y^b(z):) 
\, ,\nn
 \eea
 The  mass of the string states is
  \begin{eqnarray}
   M^2=\frac12 m^{2}_{L}+ \frac12m^{2}_{R}&=&\frac12 k_{aL}.k_{aL}+\frac12 
k_{aR}.k_{aR}+2 ( N+\bar N-2)
\label{stringmasses}
  \end{eqnarray}
 where $  N, \bar N$ are the number of string oscillators and the level 
matching condition reads 
\bea
\frac14 k_{aL}.k_{aL}-\frac14 
k_{aR}.k_{aR}- (N-\bar N)&=& p_n\tilde p^n-(N-\bar N)= 0
\eea
 
and similarly for the right moving one.
\subsection{Torus example}
The frame base can be written as  (as mention the factor $\sqrt2$ is included 
to maintain the normalization conditions $\alpha^2=2$ for simple roots) 
 \bea 
e_{1}&=&\frac{1}{\sqrt2}( 0, {\sqrt{G_{11}}} ), \qquad   e_{2} =
\frac{1}{\sqrt2}( 
\frac{\sqrt{detG}}{\sqrt{G_{11}}},  \frac{G_{12}}{\sqrt{G_{11}}} 
),\qquad \\\nn 
\eea
leading to the  matrix $g_{mn}=e_{m}.e_{n}=\frac12 G_{mn}$. 
with dual lattice vectors ($e_{m}^{*}=e^{m}$)
\bea 
e_{1}^{*}& =& \sqrt2(  -\frac{G_{12}}{\sqrt{detG}\sqrt 
{G_{11}}},\frac{1}{\sqrt{G_{11}}} ), \qquad e_{2}^{*}= \sqrt2( \frac{\sqrt 
G_{11}}{\sqrt{detG}},0 ) \nn
\eea

A field $ 
  B=
B_{12}\left(\begin{matrix}
    0&1\\
    1& 0
  \end{matrix}\right)
$ can also be introduced.  When 
\bea
G=\left(\begin{matrix}
    2& -1\\
    -1& 2
  \end{matrix}\right)
\eea
the $SU(3)$ Cartan matrix is obtained and frame  
vectors become $e_m=\frac{1}{\sqrt2} \alpha_m$ where
\bea
\alpha_1&=&(0,\sqrt2 ), \qquad \alpha_2= 
(\frac{\sqrt3}{\sqrt2},-\frac{1}{\sqrt2})\\
\label{klroots}
\eea
are the $SU(3)$ simple roots.

On the other hand, $G_{22}=G_{11}=2; G_{12}=0$ corresponds to an 
$SU(2)\times SU(2)$ algebra.

Metric and $B$ field define the complex structure $U  =U_1+iU_{2}$  and Khaler 
 structure $T =T_1+iT_{2}$ of the 
torus with $ U_{1}  =\frac{g_{12}}{{g_{22}}},\, U_{2} =
\frac{\sqrt{detg}}{{g_{22}}}, \, T_{1} =B_{12},\, T_{2} =\sqrt{det g}
$. In terms of complex moduli, $SU(3)_L\times SU(3)_R$ enhancing occurs at 
\bea 
T& =& -\frac12+i\frac{\sqrt3}2=U
\eea
whereas $(SU(2)\times SU(2))_L \times (SU(2)\times SU(2))_R $ enhancing is 
achieved for 
\bea 
T& =& i =U
\eea

\section{General enhancing groups} 
\label{sec:General enhancing groups} 
 We show here that,  in the general case of an 
enhancing from $U(1)^{r}_{L}\times U(1)^{r}_{R}$ to a gauge group $G_{L}\times 
G_{R}$ the generalized fluxes lead to the  the exact vector and scalar  massive 
terms. Namely, the corresponding masses coincide with the masses computed from 
string theory.
Consider 
the L-R splitting of indices in the $\cal C$ base $A=(a,\hat{a})$ where the 
first (second)  entries belong to left group $G_{L}$ (right group 
$G_R$). Let us focus on $G_L$ and further split left indices as 
$a=(\alpha,I)$  corresponding to charged generators and  Cartan generators 
$I=1,\dots r$ (and similarly for Right group).
\subsection{Vector masses} 
The vector mass terms in the Lagrangian read 
 \begin{eqnarray}\nn
  \big( f_{ABC} A^{B}_{\mu} M^{C}_{D} + f_{DBC} A^{B}_{\mu}M^{C}_{A}  
\big)^{2}&\sim & A^{B}_{\mu}A^{E\,\mu} f_{ABC}f_{DEF}\big( \eta^{AD}\eta^{CF} - 
\delta^{AD}\delta^{CF} \big)\\
 &\sim& A^{B}_{\mu}A^{E\,\mu}f_{aB\hat{c}}f_{aE\hat{c}}
 \end{eqnarray}
If the fluxes do not mix Left and Right sectors (as it happens at the  self 
dual point) then all vectors 
are massless. 
From momentum conservation we know that 
$f_{a I \hat{c}}=f_{a\bar I \hat{c}}=0$. Moreover $a$ and 
$\hat{c}$ can not be charged indices simultaneously. Then
 \begin{equation}
 f_{a B\hat {c}}f_{a E\hat {c}} = 
f_{I B\hat \gamma }f_{I E\hat \gamma} 
+ f_{\alpha B\hat I}f_{\alpha E\hat I} 
 \end{equation}

  We conclude that indices $B,E$ in the previous expression must be 
charged indices and, moreover, they must be equal by momentum conservation
  \begin{equation}
  \begin{aligned}
  A^{B}_{\mu}A^{E\,\mu} f_{aB\hat {c}}f_{aE\hat {c}}&\sim 
\sum_{\hat \gamma} 
A^{\hat \gamma}_{\mu}A^{\hat \gamma\,\mu}\sum_{I=1}^{r} 
f_{I\,-\hat \gamma\,\hat \gamma}f_{I\,-\hat \gamma\,\hat \gamma}
+ \sum_{\alpha} A^{\alpha}_{\mu}A^{\alpha\,\mu} \sum_{\hat I=1}^{r} 
f_{\alpha\,-\alpha\hat I}f_{\alpha\,-\alpha\hat I}\\
&= \sum_{\hat \gamma} A^{\hat \gamma}_{\mu}A^{\hat \gamma\,\mu} 
m^{2}_{\hat \gamma} + \sum_{\alpha}A^{\alpha}_{\mu}A^{\alpha\,\mu} 
m^{2}_{\alpha}
 \end{aligned}
 \label{massvb}
  \end{equation}
  where the sum runs over the positive roots. By using that (see   
(\ref{structureconstvertices}))
$f_{I\,-\hat \gamma\,\hat \gamma}=K_{L, \,\,\hat \gamma}^{I}$ i.e. 
the 
I-component of the Left $\gamma$ momentum   (similar for the right case) we 
can write the masses as  
$m^{2}_{\hat \gamma}=\sum_{I=1}^{r} (K_{L, \,\,\hat \gamma}^{I})^{2} $ and 
for the $\gamma$-left vector is $m^{2}_{\gamma}=\sum_{I=1}^{r} (K_{R, 
\,\,\hat \gamma}^{I})^{2} $, that coincide with vector masses computed from 
\eqref{stringmasses}.
\subsection{Scalar masses}
We denote the, $(dimG-r)^2$, massless scalars charged under Left and Right  
gauge group as 
$M^{\alpha \hat \beta}$. In string compactification they are described by 
the vertex operators $V^{\alpha \hat \beta}(z,\bar z)\propto J^{\alpha}(z){\hat 
J}^{\hat \beta}(\bar z)$ with $J^{\alpha}(z)= e^{k_{L{\alpha}}.y}$. When moving 
away from the self-dual point a non vanishing Right contribution 
$k_{R{\alpha}}$ ($m_-$ in circle example) appears and similarly a $k_{L{\hat 
\beta}}$, from the Right sector.
Therefore, the scalar Left and Right internal momenta become
\begin{eqnarray}
 k_{L{\alpha \hat \beta}}&=& k_{L{\alpha}}+k_{L{\hat \beta}}\\\nn
 k_{R{\hat \beta \alpha }}&=& k_{R{\hat \beta}}+k_{R{\alpha}}\\\nn
 \label{kscalarvector}
\end{eqnarray}
Recall that, since $N=\bar N=0$ level matching requires $ k_{L{\alpha \hat 
\beta}}^2= k_{R{\hat \beta \alpha 
}}^2$.

The mass of the scalar is \eqref{stringmasses}
\begin{equation}
 M_{\alpha \hat \beta}^2= \frac12 k_{L{\alpha \hat \beta}}^2+ \frac12  k_{R{\hat 
\beta \alpha }}^2-4
\end{equation}
By replacing the values \eqref{kscalarvector} into this formula and by 
using LMC for vector currents 
\begin{eqnarray}
 k_{L{\alpha}}^2-k_{R{\alpha}}^2&=&1\\\nn
 k_{L{\hat \beta  
}}^2- k_{R{\hat \beta }}^2&=&-1
\end{eqnarray}
we obtain
\begin{equation}
 M_{\alpha \hat \beta}^2= k_{R{\alpha}}.(k_{R{\alpha}}+ k_{R{\hat \beta }})+
 k_{L{\hat \beta }}.( k_{L{\hat \beta }}+ k_{L{\alpha }})
\end{equation}
that, as expected, vanishes at the fixed point. 
By using the identification 
with fluxes \eqref{structureconstvertices} this expression can be recast as
\begin{equation}
 M_{\alpha \hat \beta}^2= f_{\hat I \alpha -\alpha}(  f_{\hat I \alpha -\alpha}+
  f_{\hat I \hat \beta -\hat \beta})+
    f_{I \hat \beta -\hat \beta}(f_{I \hat \beta -\hat \beta} +f_{ I \alpha 
-\alpha})
\end{equation}
This is exactly the combination of fluxes that appears in front of the 
quadratic scalar term when we mimmick the steps we followed for the circle 
case\eqref{su2scalarmass}.
Namely,  insert  the expansion in  
scalar fluctuations $M$ \eqref{Hintc}, 
into the third row of the DFT action \eqref{dftaction} and use the values 
\eqref{structureconstantsr}.


\begin{thebibliography}{98}

\bibitem{Narain:1985jj}
  K.~S.~Narain,
  ``New Heterotic String Theories in Uncompactified Dimensions < 10,''
  Phys.\ Lett.\  {\bf 169B} (1986) 41.
  doi:10.1016/0370-2693(86)90682-9
\bibitem{Giveon:1994fu}
  A.~Giveon, M.~Porrati and E.~Rabinovici,
  ``Target space duality in string theory,''
  Phys.\ Rept.\  {\bf 244} (1994) 77
  [hep-th/9401139].

\bibitem{agimnr}
G.~Aldazabal, M.~Grana, S.~Iguri, M.~Mayo, C.~Nu\~nez and J.~A.~Rosabal,
  ``Enhanced gauge symmetry and winding modes in Double Field Theory,''
  JHEP {\bf 1603}, 093 (2016)
  [arXiv:1510.07644 [hep-th]].
  
  \bibitem{reviewamn}
G.~Aldazabal, D.~Marqu\'es and C.~Nu\~nez,
  ``Double Field Theory: A Pedagogical Review,''
  Class.\ Quant.\ Grav.\  {\bf 30} (2013) 163001
  [arXiv:1305.1907 [hep-th]].
  
  \bibitem{effective}
    G.~Aldazabal, W.~Baron, D.~Marqu\'es and C.~Nu\~nez,
  ``The effective action of Double Field Theory,''
  JHEP {\bf 1111}, 052 (2011)
  [arXiv:1109.0290 [hep-th]].\\
  D.~Geissbuhler,
 ``Double Field Theory and N=4 Gauged Supergravity,''
  JHEP {\bf 1111}, 116 (2011)
  [arXiv:1109.4280 [hep-th]].
  
    
  \bibitem{edft}
   D.~Geissbuhler, D.~Marques, C.~Nu\~nez and V.~Penas,
  ``Exploring Double Field Theory,''
  JHEP {\bf 1306}, 101 (2013)
  [arXiv:1304.1472 [hep-th]].
  
\bibitem{Hitchin}
 N.~Hitchin,
   ``Generalized Calabi-Yau manifolds,''
   Quart.\ J.\ Math.\ Oxford Ser.\  {\bf 54} (2003) 281
   [arXiv:math.dg/0209099].\\
   M.~Gualtieri, ``Generalized Complex Geometry,''
   Oxford University DPhil thesis (2004) [arXiv:math.DG/0401221].


\bibitem{GMPW}
M.~Grana, R.~Minasian, M.~Petrini and D.~Waldram,
  ``T-duality, Generalized Geometry and Non-Geometric Backgrounds,''
  JHEP {\bf 0904}, 075 (2009)
  [arXiv:0807.4527 [hep-th]].


 \bibitem{WaldramOdd}
 A.~Coimbra, C.~Strickland-Constable and D.~Waldram,
  ``Supergravity as Generalised Geometry I: Type II Theories,''
  JHEP {\bf 1111}, 091 (2011)
  [arXiv:1107.1733 [hep-th]].

\bibitem{hz}
  C.~Hull and B.~Zwiebach,``Double Field Theory,''\\
  JHEP {\bf 0909}, 099 (2009) 
  [arXiv:0904.4664 [hep-th]].

  O.~Hohm, C.~Hull and B.~Zwiebach,
  ``Background independent action for double field theory,''
  JHEP {\bf 1007}, 016 (2010)
  [arXiv:1003.5027 [hep-th]].

  O.~Hohm, C.~Hull and B.~Zwiebach,
  ``Generalized metric formulation of double field theory,''
  JHEP {\bf 1008}, 008 (2010)
  [arXiv:1006.4823 [hep-th]].

\bibitem{duff}
M.~J.~Duff,
  ``Duality Rotations In String Theory,''
  Nucl.\ Phys.\ B {\bf 335} (1990) 610.

   M.~J.~Duff and J.~X.~Lu,
  ``Duality Rotations In Membrane Theory,''
  Nucl.\ Phys.\ B {\bf 347} (1990) 394.
  
 \bibitem{tseytlin} 
   A.~A.~Tseytlin,
  ``Duality Symmetric Formulation of String World Sheet Dynamics,''
   Phys.\ Lett.\ B {\bf 242}, 163 (1990).

    A.~A.~Tseytlin,
  ``Duality symmetric closed string theory and interacting chiral scalars,''
    Nucl.\ Phys.\ B {\bf 350}, 395 (1991).

    \bibitem{siegel}
  W.~Siegel,
  ``Superspace duality in low-energy superstrings,''
  Phys.\ Rev.\ D {\bf 48} (1993) 2826
  [hep-th/9305073].

  W.~Siegel,
  ``Two vierbein formalism for string inspired axionic gravity,''
  Phys.\ Rev.\ D {\bf 47} (1993) 5453
  [hep-th/9302036].


 \bibitem{hk} O.~Hohm and S.~K.~Kwak,
  ``Double Field Theory Formulation of Heterotic Strings,''
  JHEP {\bf 1106}, 096 (2011)
  [arXiv:1103.2136 [hep-th]].
  

\bibitem{Jeon:2011cn} I.Jeon, K.Lee and J. H.Park,
``Stringy differential geometry, beyond 
Riemann'',  Phys.\ Rev.\ D {\bf 84} (2011) 044022
  doi:10.1103/PhysRevD.84.044022  [arXiv:1105.6294 [hep-th]].


\bibitem{berwest}
 D.~S.~Berman and D.~C.~Thompson,
  ``Duality Symmetric String and M-Theory,''
  arXiv:1306.2643 [hep-th].


\bibitem{reviews}
  O.~Hohm, D.~Lust and B.~Zwiebach,
  ``The Spacetime of Double Field Theory: Review, Remarks, and Outlook,''
  Fortsch.\ Phys.\  {\bf 61}, 926 (2013)
  [arXiv:1309.2977 [hep-th]].

  \bibitem{alphap}
  O.~Hohm, W.~Siegel and B.~Zwiebach,
  ``Doubled $\alpha'$-geometry,''
  JHEP {\bf 1402}, 065 (2014)
  [arXiv:1306.2970 [hep-th]].

       O.~Hohm and B.~Zwiebach,
  ``Double field theory at order $\alpha'$,''
  JHEP {\bf 1411}, 075 (2014)
  [arXiv:1407.3803 [hep-th]].

    O.~A.~Bedoya, D.~Marques and C.~Nu\~nez,
  ``Heterotic $\alpha$'-corrections in Double Field Theory,''
  JHEP {\bf 1412}, 074 (2014)
  [arXiv:1407.0365 [hep-th]].

      D.~Marques and C.~A.~Nu\~nez,
  ``T-duality and $\alpha'$-corrections,'' JHEP {\bf 1510}, 084 (2015)
     [arXiv:1507.00652 [hep-th]].



\bibitem{alphap2}
 A.~Coimbra, R.~Minasian, H.~Triendl and D.~Waldram,
  ``Generalised geometry for string corrections,''
    arXiv:1407.7542 [hep-th].

  \bibitem{ss}
  J.~Scherk and J.~H.~Schwarz,
  ``How to Get Masses from Extra Dimensions,''
  Nucl.\ Phys.\ B {\bf 153}, 61 (1979).

\bibitem{Samtleben:2008pe}
  H.~Samtleben,
  ``Lectures on Gauged Supergravity and Flux Compactifications,''
  Class.\ Quant.\ Grav.\  {\bf 25} (2008) 214002
  [arXiv:0808.4076 [hep-th]].
\bibitem{Trigiante:2016mnt}
  M.~Trigiante,
  ``Gauged Supergravities,''
  arXiv:1609.09745 [hep-th].
\bibitem{gm} M.~Grana and D.~Marques,
  ``Gauged Double Field Theory,''
  JHEP {\bf 1204}, 020 (2012)
  [arXiv:1201.2924 [hep-th]].


\bibitem{Aldazabal:2016yih}
  G.~Aldazabal, M.~Mayo and C.~Nu\~nez,
  ``Probing the String Winding Sector,''
  JHEP {\bf 1703} (2017) 096
  [arXiv:1611.04927 [hep-th]].

\bibitem{dmr}
 G.~Dibitetto, J.~J.~Fernandez-Melgarejo, D.~Marques and D.~Roest,
  ``Duality orbits of non-geometric fluxes,''
  Fortsch.\ Phys.\  {\bf 60}, 1123 (2012)
  [arXiv:1203.6562 [hep-th]].



  \bibitem{waldramgaugings}
  K.~Lee, C.~Strickland-Constable and D.~Waldram,
  ``New gaugings and non-geometry,''
  arXiv:1506.03457 [hep-th].

\bibitem{Blumenhagen:2014gva}
  R.~Blumenhagen, F.~Hassler and D.~Lüst,
  ``Double Field Theory on Group Manifolds,''
  JHEP {\bf 1502} (2015) 001
  doi:10.1007/JHEP02(2015)001
  [arXiv:1410.6374 [hep-th]].
HEP :: Buscar ::  Ayuda ::  Terms of use ::  Privacy policy 
Powered by Invenio v1.1.2+ 
Problems/Questions to feedback@inspirehep.net 


  \bibitem{schulz}
  M.~B.~Schulz,
  ``T-folds, doubled geometry, and the SU(2) WZW model,''
  JHEP {\bf 1206}, 158 (2012)
  [arXiv:1106.6291 [hep-th]].


\bibitem{Lee:2014mla}
  K.~Lee, C.~Strickland-Constable and D.~Waldram,
  ``Spheres, generalised parallelisability and consistent truncations,''
  arXiv:1401.3360 [hep-th].


  \bibitem{Dall}
  G.~Dall'Agata, G.~Inverso and M.~Trigiante,
  ``Evidence for a family of SO(8) gauged supergravity theories,''
  Phys.\ Rev.\ Lett.\  {\bf 109}, 201301 (2012)
  [arXiv:1209.0760 [hep-th]].


 
\bibitem{DFThettype2}
  O.~Hohm and S.~K.~Kwak,
  ``Double Field Theory Formulation of Heterotic Strings,''
  JHEP {\bf 1106} (2011) 096
  [arXiv:1103.2136 [hep-th]].


\bibitem{eft}
  O.~Hohm and H.~Samtleben,
  ``Gauge theory of Kaluza-Klein and winding modes,''
  Phys.\ Rev.\ D {\bf 88} (2013) 085005
  [arXiv:1307.0039 [hep-th]].\\ 
  G.~Aldazabal, M.~Gra\~na, D.~Marqu\'es and J.~A.~Rosabal,
  ``The gauge structure of Exceptional Field Theories and the tensor 
hierarchy,''
  JHEP {\bf 1404} (2014) 049
  [arXiv:1312.4549 [hep-th]].

\bibitem{Cagnacci:2017ulc}
  Y.~Cagnacci, M.~Gra\~na, S.~Iguri and C.~Nú\~nez,
  ``The bosonic string on string-size tori from double field theory,''
  arXiv:1704.04242 [hep-th].

\end{thebibliography}
 \end{document}